\documentclass[10pt,twocolumn]{revtex4-2} 
\usepackage[dvips]{color}
\usepackage{epsfig}
\usepackage{amssymb}
\usepackage{latexsym}
\usepackage[dvipdfm,truedimen,%
top=30truemm,bottom=20truemm,%
left=20truemm,right=20truemm]{geometry}
\usepackage[latin9]{inputenc}
\usepackage{graphicx}
\def\itmb{\begin{itemize}}
\def\itme{\end{itemize}}
\def\enmb{\begin{enumerate}}
\def\enme{\end{enumerate}}
\def\eqnb{\begin{equation}}
\def\eqne{\end{equation}}

\makeatother

\draft 

\begin{document}


\title{ Supersymmetry in Hadron Spectroscopy and \\ Solitons in 2 Dimensional Fermionic Media
} 

\author{ Sadataka Furui}
\affiliation{(Formerly) Graduate School of Science and Engineering, Teikyo University; furui@umb.teikyo-u.ac.jp}
%
\date{\today}
\begin{abstract}
An overview of hadron spectroscopy obtained from an extension of superconformal field theory of de Alfaro, Fubini and Furlan, which originates from the supergauge transformation of Wess and Zumino and
influenced on the string theory are preseted.
  
  The standard model of hadron dynamics is based on gauge theory. 
In the 4 dimensional ($4D$) QCD Lagrangian, the Faddeev-Popov ghost that appears to compensate unphysical degrees of freedom in internal fermion propagators can be regarded as the partner of a fermion. 
Donaldson formulated a $5D$ gauge theory. 

Fubini and Rabinovici formulated superconformal quantum mechanics in de Sitter space, and Brodsky and de T\'eramond formulated relativistic light front holographic QCD (LFHQCD) in Kaluza-Klein type $5D$ anti de Sitter (AdS) space, using Dirac's light cone gauge, and succeeded in obtaining spectroscopy of baryons (fermions) and corresponding mesons (bosons), except the pion which is an axial scalar meson that remains massless.  

In the LFHQCD, the time parameter $\tau=t\pm z/c$, $(z\in {\bf C})$ appears by the choice of the light-cone gauge, or the front form of Dirac's formulation. The theory is an extension of AdS/CFT correspodence of Maldacena in which incorporation of Einstein's gravity theory in Maxwell's electromagnetic theory was tried in the framework of the string theory.

Taking mathematical frameworks of treating the supersymmetry by Brodsky, Witten, Connes and their collaborators as central thema, we summarize other supersymmetric approaches like string theory in particle physics and astrophysics. We discuss about 2 dimensional solitons whose dynamics is represented also by $\tau=t\pm z/c$, ($z\in$ pure quaternion${\bf H}$).
\end{abstract}
\maketitle 
\section{Introduction}
Supersymmetry is a symmetry between bosons and fermions. In nature, the symmetry is broken, i.e. in hadron spectroscopy, spectra of baryons which are fermions do not always have corresponding spectra of mesons which are bosons. Similarly the ground state of the standard model consists of a boson and there does not exists a fermion partner.

 Spectroscopies are defind in Hilbert space. In the formulation of  Wess-Zumino (WZ) model\cite{WZ74}, a conformal structure of $\alpha(x)$ at position $x$ and metrics $\eta_{\mu\nu}$ of Lagrangian for Majorana spinors\cite{SM70, Labelle10} are defined.  The coupling constant $g$ in Lagrangian was chosen  to be positive, or interaction between Majorana spinors are repulsive and two spinors cannot have a zero distance, and the fields are conformal. 
 
 Magnetic monopoles are solitons in $3D$ space, which are confined by superconductor produced by Higgs field\cite{tHooft71,tHooft74,tHooft06}. `t Hooft discussed open string theory by transforming magnetic monopoles which can be regarded as 3 dimensional solitons to quarks, and possibility of describing hadrons. 
 We do not consider solitons in $3D$ space in this review, but explain the string theory and study propagation of solitons in $2D$ space that appear by nonlinear interactions\cite{FDS20a,FDS20b}.
 
 Procedures to keep conformality by adding extra Lagrangians are discussed in the paper\cite{tHV72}. Magnetic monopoles of $SU(2)\times U(1)$ gauge group of George-Glashaw\cite{GG72} were derived by 't Hooft\cite{tHooft74}.  Instantons in $3D$ non-abelian gauge theory was found to contain a monopole solution by Polyakov\cite{Polyakov75}.  Reviews of monopole solutions are given in \cite{Polyakov87, NS83, Shnir05} as examples.
 
  In an application of gauge theory to four dimensional topology, Donaldson\cite{Donaldson83} proved an existence of a smooth ball of $5D$ with only finite number of singularities. The $r'th$ homotopy group of 
n-sphere $\pi_r(S^n)$ was systematically studied by Rohlin, whose theory is summarized in\cite{GM86}, and Donaldson extended Rohlin's theory. For homotopy groups on Riemann surface, please refer to \cite{BT82}.

The super gauge transformation that links fermions and bosons defined by WZ in space 3-dimension, 1 time dimensional system was applied to conformal quantum mechanics of space 1-dimension, 1 time dimensional system by de Alfaro, Fubini and Furlan (dAFF)\cite{dAFF76}. They defined the Hailtonian operator $H$, 
dilation generator $D$ and conformal generator $K$, that satisfy  
\[
[H,K]=-\sqrt{-1} H,  [K,D]=-\sqrt{-1} K,  [H,K]=2\sqrt{-1} D. 
\]
In dAFF, 4 components of gauge vector $A_m$ is gauge fixed to the light-cone $A_+=0$, and anti-selfdual field $F_{\mu\nu}^-$ satisfies $F_{\mu\nu}^-=-^*F_{\mu\nu}^-$, where $^*$ means Hodge dual\cite{Lounesto01,GSW86,GSW87}.

Superconformal quantum mechanics in de Sitter space was formulated by Fubini and Rabinovici \cite{FR84}. 
de Sitter space is the solution of Einstein field equation, whose coordinates satisfy 
\[
\sum_{i=1}^4(x^i)^2-(x^0)^2=R^2=\frac{3}{G\lambda}
\]
where $\lambda$ is a cosmological constant\cite{SL05}.

Maldacena\cite{Maldacena98} conjectured that our 4 dimensional world can be represented by the conformal field theory (CFT) projected on the boundary of $4+1$ dimensional anti de Sitter (AdS) space, and called the principle as AdS/CFT correspondence.  Superconformal quantum mechanics in AdS space was formulated by de T\'eramond, Brodsky and their collaborators using $5D$ holographic mapping of physical space to AdS space. For the review \cite{BdTDL16} is helpful.
They considered light cone gauge, and the light front holographic QCD, the Dirac's relativistic treatment\cite{Dirac49}.

When one extends the metrics of the Hilbert space from $4D$ Minkowsky space to $5D$ ADS space, and assumes the holographic principle, 
one can show relations between baryon spectroscopies and meson spectroscopies\cite{BdTDE15}, although the mass of lightest meson pion remains massless. 

A Higgs boson in the standard model, whose supersymmetric partner which is called higgsino is not observed experimentally. However, Higgs boson interacts with fermions and cause them to be massive, and it decays into two photons, vector boson pairs and $b\bar b$ pairs\cite{ATLAS12, CMS12}. Decay modes of a heavior Higgs boson is also proposed\cite{BGMSU16}.

Dirac \cite{Dirac63} stated that in use of mathematics of transformations of relativity and quantum mechanics, there are two ways, i) symbolic methods that deals with invariants and ii) the methods of coordinates or representations that deals with numbers corresponding to these quantities, expressed by real ${\bf R}$ numbers. He has established quantum electrodynamics (QED) based on the first method using a complex number ${\bf C}$ defining the $U(1)$ gauge transformation as an aid to practical calculations. 

Extension of standard model in ${\bf R}^{1,1}\times {\bf R}^3$ space-time in the framework of noncommutative geometry was proposed by Connes and his collaborators. Noncommutative geometry is based on K-theory in which real numbers, complex numbers and quaternions play their roles. 

Supersymmetry plays a role in fluctuation of fermionic systems (instantons) near the ground state via occurlence of bosonic soliton waves. When the algebra $\wedge({\bf R})$ is one dimensional vector space normal to a 2-dimensional plane in hysteretic solids,  a similar analysis of spectroscopy of sound wave (phonon) is possible by projecting dynamics to 2 dimensional projected quaternion space, and by taking the time parameter $\tau=t\pm z/v$ where $v$ is the sound velocity in a material.

Hysteretic effects in phonon systems and instanton effects are compared via the holonomy theory.

A Phonon can be regarded as a soliton which is a boson that propagates in fermionic media. It does not have supersymmetric partner, but its propagation depends on even or odd numbers of fermions that it interacts.

In the following subsections, WZ model, dAFF model, Witten's model, anti self-dual Yang Mills model, Penrose's model, dynamical models using quaternions are reviewed as a preparation of
hadron spectroscopy in $5D$ projective spaces and soliton dynamics in $2D$ projective quaternion space, both related to the supersymmetry.


\subsection{Wess-Zumino Model and supergauge transformation}
Wess and Zumino \cite{WZ74} defined supergauge transformation in four dimension as gauge transformation whose commutators turns out to be a combination of a conformal transformation and a $\gamma_5$ transformation.

Consider Majorana spinors $\alpha(x)$ that satisfy
\begin{eqnarray}
&&(\gamma_\mu\partial_\nu+\gamma_\nu\partial_\mu-\frac{1}{2}\eta_{\mu\nu}\gamma^\lambda\partial_\lambda)
\left(\begin{array}{c}
\alpha_1\\
\alpha_2 \end{array}\right)=0,
\end{eqnarray}
where
\begin{equation}
\gamma^0=\gamma_0=\left(\begin{array}{cc}
0 & I\\
I & 0 \end{array}\right),\quad\gamma^k=-\gamma_k=\left(\begin{array}{cc}
0 & -\sigma_k\\
\sigma_k & 0 \end{array}\right), \nonumber
\end{equation}
where $k=1,2,3$ and
\begin{eqnarray}
&&\{\gamma^\mu,\gamma^\nu\}=\gamma^\mu\gamma^\nu+\gamma^\nu\gamma^\mu=2g^{\mu\nu}{\bf I},\nonumber\\
&&\gamma^5=i\gamma^0\gamma^1\gamma^2\gamma^3=\gamma_5=\left(\begin{array}{cc}
                                                                            I& 0\\
                                                                            0&- I\end{array}\right),\nonumber\\
&&\sigma_1=\left(\begin{array}{cc}
                           0&1\\
                           1&0\end{array}\right), \sigma_2=\left(\begin{array}{cc}
                           0&-\sqrt{-1}\\
                           \sqrt{-1}&0\end{array}\right),\nonumber\\
 &&\sigma_3=\left(\begin{array}{cc}
                           1&0\\
                           0&-1\end{array}\right).
\end{eqnarray}
The infinitesimal parameters $\xi_\mu$ satisfy
\begin{equation}
\partial_\mu\xi_\nu+\partial_\nu\xi_\mu=\frac{1}{2}\eta_{\mu\nu}\gamma^\lambda\partial_\lambda.
\end{equation}
Then it follows that
\begin{equation}
\xi_\mu=2\sqrt{-1}\bar\alpha_1\gamma_\mu\alpha_2,\nonumber
\end{equation}
and by multiplying $\gamma^\mu$ to (2), one obtains
\begin{equation}
\partial_\mu\alpha=\frac{1}{4}\gamma_\mu\gamma^\lambda\partial_\lambda\alpha.\nonumber
\end{equation}
It was shown that $\alpha$ can be expressed in the form
\begin{equation}
\alpha=\left(\begin{array}{c}\alpha^{(0)}_1\\
                                   \alpha^{(0)}_2\end{array}\right)+
                                   \gamma_\mu x^\mu \left(\begin{array}{c} \alpha^{(1)}_1\\
                                                                                      \alpha^{(1)}_2\end{array}\right),\nonumber
\end{equation}
and
\begin{equation}
\xi_\mu=c_\mu+\omega_{\mu\nu}x^\nu+\epsilon x_\mu+a_\mu x^2-2x_\mu \alpha\cdot x.\nonumber
\end{equation}
The metric tensor becomes
\begin{equation}
\eta=4\sqrt{-1}(\bar\alpha^{(1)}_1\gamma_5 \alpha^{(0)}_2-\bar\alpha^{(1)}_2\gamma_5\alpha^{(0)}_1),\nonumber
\end{equation}
which mixes $\bar\alpha\psi$ and $\alpha\gamma_5\psi$.  

The supergauge transformation was applied to spin systems with superfield $\phi(t,\theta,\bar\theta)$ where $\theta, \bar\theta$ are totally anti-commuting quanities and $t\in {\bf R}$ by Nicolai \cite{Nicolai76}, 
\begin{equation}
\phi(t,\theta,\bar\theta)=A(t)+\theta\psi(t)+\bar\psi(t)\bar\theta+\theta\bar\theta F(t),\nonumber
\end{equation}
where $A$ is a fermionic function  that satisfies $A^*=A$, $\psi_j\psi_k$, $\psi_i \psi_j \psi_k$ etc. are fermi fields, $F(t)$ is a bosonic function that satisfies $F^*=F$. 
Similar model was constructed also by Polyakov\cite{Polyakov87}.

The Hamiltonian $H=\{ Q, Q^\dagger\}$ where $Q=\sum_{j\in G}\sum_{\hat n} a_j^*\psi_{j+l\hat n}$, $Q^\dagger=\sum_{j\in G}\sum_{\hat n} \bar\psi_{j+l\hat n}a_j$, where $G$ are lattice points and $\hat n$ are unit vectors at lattice points.

The function of superfields $\phi(t,\theta,\bar\theta)$, satisfies
\begin{equation}
[H,\phi]=\sqrt{-1}\frac {d}{dt}\phi.\nonumber
\end{equation}

A review of super gauge transformation is also given by Green et al \cite{GSW86,GSW87}.

\subsection{de Alfaro-Fubini-Furlan model}
 de Alfaro, Fubini and Furlan (dAFF)\cite{dAFF76} studied a field theory in one over-all time dimension, invariant under the full conformal group, whose Lagrangian of a physical operator $Q(t)$ with a coupling constant $g>0$ is given by
\begin{equation}
L=\frac{1}{2}({\dot Q}^2-\frac{g}{Q^2}).
\nonumber
\end{equation}
They considered fundamental operators: 1) $H=\frac{1}{2}p^2+\frac{g}{Q^2}$, corresponding to hamiltonian\cite{Dirac49,Dirac50}, 2) $D$ corresponding to dilatation operation, and 3) $K$ corresponding to conformal transformation. They constructed $O(2,1)$ conformal group of non-compact rotations
\begin{equation}
R=\frac{1}{2}(\frac{1}{a}K+a H),\quad S=\frac{1}{2}(\frac{1}{a}K-a H),
\nonumber
\end{equation}
where $a$ is a constant with dimension of length. Although the lowest eigenstate $H$ is not normalizable due to infrared divergence, eigenstates of $R$ are meaningful. Combinations
\begin{equation}
G=u\, H+v\,D+w\,K \nonumber
\end{equation}
satisfies the evolution equation
\begin{equation}
\frac{\partial G}{\partial t}=\sqrt{-1}[H,G]=0,\nonumber
\end{equation}
and rotations become compact when $\Delta=v^2-4u\,w>0$.

\subsection{Witten's model and Fubini-Rabinovici model in de Sitter space}
In nature supersymmetry is broken, and constraints on supersymmetry breaking in the Yang-Mills field \cite{YM54} was discussed by Witten\cite{Witten82a,Witten82b}.  In zero momentum space, supersymmetry charge $Q_1,\cdots,Q_k$ are chosen to satisfy
\begin{eqnarray}
&&Q_1^2=Q_2^2=\cdots =Q_k^2=H,\nonumber\\
&&Q_i Q_j+Q_j Q_i=0,\quad {\rm for}\quad i\ne j.
\end{eqnarray}

Bosonic states are defined by 
\begin{equation}
exp(2\pi \sqrt{-1} J_z)|b\rangle=|b\rangle
\nonumber
\end{equation}
and fermionic states are defined by
\begin{equation}
exp(2\pi \sqrt{-1} J_z)|f\rangle=-|f\rangle,\nonumber
\end{equation}
where $J_z$ is a rotation operator, and the operator 
\begin{equation}
(-1)^F=exp(2\pi\sqrt{-1}J_z)\nonumber
\end{equation}
distinguishes fermions with half integer $J_z$ and bosons with integer $J_z$.  For a particle moving in one dimension, he considered supersymmetric charge
\begin{eqnarray}
Q_1&=&\frac{1}{2}(\sigma_1 p+\sigma_2 W(x)), \nonumber\\
Q_2&=&\frac{1}{2}(\sigma_1 p-\sigma_2 W(x)),
\end{eqnarray}
where $W(x)$ is an arbitraly function.

The number of zero energy states of bosons $n_B^{E=0}$ and that of fermions $n_F^{E=0}$ have the relation
\[
Tr (-1)^F=n_B^{E=0}-n_F^{E=0},
\]
and when $n_B^{E=0}-n_F^{E=0}=0$, there are two possibilities, 1) $n_B^{E=0}=n_F^{E=0}=0$,  supersymetry is broken, 2) $n_B^{E=0}=n_F^{E=0}$ are equal but non zero, supersymmetry is unbroken.

When $n_B^{E=0}-n_F^{E=0}\ne 0$, supersymmetry is not spontaneously broken. Spontaneous symmetry breaking occurs in the Higgs mechanism.

Fubini and Rabinovici \cite{FR84} extended the operator $Q(t)$ as
\begin{eqnarray}
Q&=&\sum_{\alpha=1}^D\psi_\alpha^+(-\sqrt{-1}p_\alpha+\frac{dW}{dx_\alpha}),\nonumber\\
Q^+&=&\sum_{\alpha=1}^D\psi(\sqrt{-1}p_\alpha+\frac{dW}{dx_\alpha}),
\end{eqnarray}
where $W(x)$ is the superpotential and $p_\alpha=-\sqrt{-1}(\partial/\partial x_\alpha)$. The ${\mathcal N}=1$ supersymmetry was extended to ${\mathcal N}=2$ de Sitter supersymmetry. 

The algebra of dAFF\cite{dAFF76} was extended by defining
\[
G=u H+ v D+ w K, \quad \Delta=v^2-4 u w<0,
\]
and incorpolating Witten's algebra
\[
\frac{1}{2}\{Q,Q^+\}=H, \quad \{Q,Q\}=\{Q^+, Q^+\}=0.
\] 

Here 
\[
Q=\psi^+(-\sqrt{-1}p+\frac{dW}{dx}),\quad Q^+=\psi(\sqrt{-1}p+\frac{dW}{dx}) 
\]
and super potential $W(x)=\frac{1}{2} f \log x^2$. An extension to spacially many dimensions and system with number of fermions and bosons doubled ${\mathcal N}=2$ and transitions from strongly coupled vacuum to weakly coupled monopole condensates following the works of Bogomol'nyi \cite{Bogomolnyi76}, Prasad and Sommerfield \cite{PS75} were presented by Seiberg and Witten\cite{SW94}.

\subsection{Anti self-dual Yang-Mills  (ASDYM) field model}
In the anti self-dual Yang-Mills  (ASDYM) equation framework\cite{DW07,Dunajski10} based on the twistor theory \cite{AW77, Ward81, PR86}, one can construct integrable $SU(2)$ chiral model, including the Higgs field $\Phi$\cite{EB64,Higgs66,Kibble67} and a vector field $A$
\begin{eqnarray}
{\hat \Phi}&=&\frac{\Phi}{|\Phi|},\nonumber\\
A_i^a&=&-\epsilon^{abc}\partial_i{\hat \Phi}^b {\hat \Phi}^c+k_i {\hat \Phi}^a, 
\end{eqnarray}
for some vector $k_i$.
The magnetic field is $B_k=\frac{1}{2}\epsilon_{ijk}F_{ij}^a{\hat\Phi}^a$.

In Minkowski space defined by $ds^2=-dt^2+dx^2+dy^2$, the Lagrangian density
\begin{eqnarray}
{\mathcal L}&=&\frac{1}{2}Tr(F_{\mu\nu}F^{\mu\nu})-Tr(D_\mu\Phi D^\mu\Phi),\nonumber\\
F_{ij}^a&=&\partial_i A_j^a-\partial A_i^a-\epsilon_{abc}A_i^b A_j^c\nonumber\\
&=&2\epsilon^{abc}\partial_i{\hat\Phi}^b\partial_j {\hat \Phi}^c-(\epsilon^{pqr}\partial_i{\hat\Phi}^p\partial_j {\hat \Phi}^q) {\hat \Phi}^a,\nonumber\\
D_x\Phi&=&F_{yt},\quad D_y\Phi=F_{tx},\quad D_t\Phi=F_{yx}.
\end{eqnarray}

Solutions of the equation
\begin{equation}
\partial_\mu K_\nu-\partial_\nu K_\mu=\frac{1}{4}\eta_{\mu\nu}\partial_\rho K^\rho\nonumber
\end{equation}
is called Killing vectors and one parametrizes $K=\partial/\partial\tau$
and transform real coordinates $(t,x,y,\tau)$ to
\begin{equation}
z=\frac{x-\tau}{\sqrt 2}, \quad \tilde z=\frac{x+\tau}{\sqrt 2},\quad w=\frac{t+y}{\sqrt 2},\quad \tilde w=\frac{t-y}{\sqrt 2}.\nonumber
\end{equation}
Using the Hodge operator $*$ on ${\bf R}^{2,1}$, the Lax pair equation can be written as
\begin{equation}
D\Phi=*F. 
\end{equation}
For $GL(2, {\bf R})$ valued functions $A_x, A_y$ which depend on $(x,y)$, one defines a two component vector $v$,
which satisfies
\begin{equation}
D_x v=\partial_x v+A_x v=0, \quad D_y v=\partial_y v+A_y v=0,\nonumber
\end{equation}
and 
\begin{eqnarray}
&&\partial_y\partial_x v-\partial_x\partial_y v=-\partial_y(A_x v)+\partial_x(A_y v)\nonumber\\
&&=(\partial_xA_y-\partial_y A_x+[A_x,A_y])v=0.\nonumber
\end{eqnarray}
The above equation is equivalent to
\begin{equation}
F_{xy}=[D_x,D_y]=0\nonumber
\end{equation}
When one considers with a complex projective parameter $\lambda\in {\bf CP}^1$,
\begin{equation}
L=D_{\bar z}-\lambda D_w, \quad M=D_{\bar w}-\lambda D_z,\nonumber
\end{equation}
ASDYM becomes
\begin{equation}
[L,M]=F_{\bar z\bar w}-\lambda(F_{w\bar w}-F_{z\bar z})+\lambda^2F_{wz}=0.\nonumber
\end{equation}

The Lax pair equations for $GL(n,{\bf C})$ valued functions $\Psi(w,z,\tilde w,\tilde z,\lambda)$, where $w,z,\tilde w,\tilde z\in {\bf R}$, $\lambda\in{\bf C}$, satisfy
\begin{eqnarray}
L_0\Psi&=&(D_y+D_t-\lambda(D_x+\Phi))\Psi=0,\nonumber\\
L_1\Psi&=&(D_x-\Phi-\lambda(D_t-D_y))\Psi=0,
\end{eqnarray}
and
\begin{equation}
\Psi(x^\mu,\bar\lambda)^*\Psi(x^\mu,\lambda)=1.\nonumber
\end{equation}

The function $\Psi$ can be chosen to satisfy gauge transformations $J:{\bf R}^3\to U(n)$ such that
\begin{equation}
A_t=A_y=\frac{1}{2}J^{-1}(J_t+J_y),\quad A_x=-\Phi=\frac{1}{2}J^{-1}J_x,
\label{f1}
\end{equation}
and the equation (\ref{f1}) becomes 
\begin{equation}
(J^{-1}J_t)_t-(J^{-1}J_x)_x-(J^{-1}J_y)_y-[J^{-1}J_t, J^{-1}J_y]=0.\nonumber
\end{equation} 
The last term is called the Wess-Zumino-Witten (WZW) term. The Lax pair have soliton solutions\cite{Ward88}.


\subsection{Twistors and conformal model}
 In order to combine the electro magnetic force described by Maxwell equation and the gravity described by Einstein equation, Penrose proposed a twistor method\cite{PR86, Dunajski10} using a function $f$ of ${\bf R}^3\to{\bf R}$
\begin{eqnarray}
 &&v(x,y,\zeta ,t)=\nonumber\\
 &&\frac{1}{2\pi \sqrt{-1}} \oint_{\Gamma\subset {\bf CP}^1} f(-(x+\sqrt{-1}y)+\lambda(t-\zeta),\nonumber\\
 &&\quad\quad\quad\quad  (t+\zeta)+\lambda(-x+\sqrt{-1}y),\lambda )d\lambda\nonumber
\end{eqnarray}

The integral is along a curve on a manifold ${\bf CP}^1$, which is the quotient of ${\bf C}^2\times {\bf C}^2$ by the equivalence relation
\begin{equation}
({\bf Z}^0,{\bf Z}^1)\sim (c{\bf Z}^0, c{\bf Z}^1)\nonumber
\end{equation}
for $c\in{\bf C}$, and $\lambda={\bf Z}^1/{\bf Z}^0$, $\bar\lambda=1/\lambda$.

Mapping between local coordinate between $S^2$ and ${\bf CP}^1$ is given by\cite{Dunajski10}
\begin{equation}
(u_1,u_2,u_3)\to \{1-u_3, u_1+\sqrt{-1} u_2\},\nonumber
\end{equation}
and by patching two $S^2$, one obtains the Riemann sphere $S^3$.

On the manifold $S^3\times S^1$, a Lie group $GL(2, {\bf C})$ operates \cite{Kodaira66}.

\subsection{Dynamics on ${S}^3$ manifold and quaternions ${\bf H}$} 
In quantum chromo dynamics (QCD), there is axial gauge transformation $U(1)_A$, whose symmetry is spontaneously broken. For practical calculations, numbers other than complex numbers could be useful. In 1877, Frobenius showed that there are only three isomorphically distinct real finite dimensional associative division algebra ${\bf R, C}$ and quaternion ${\bf H}$.  Quaternions have bases
\begin{eqnarray}
&&{\bf I}=\left(\begin{array}{cc}
1& 0\\
0 & 1\end{array}\right),\quad 
{\bf i}=\left(\begin{array}{cc}
0& \sqrt{-1}\\
\sqrt{-1} & 0\end{array}\right)\nonumber\\
&&{\bf j}=\left(\begin{array}{cc}
0& 1\\
-1 & 0\end{array}\right),\quad
{\bf k}=\left(\begin{array}{cc}
\sqrt{-1}& 0\\
0 & -\sqrt{-1}\end{array}\right).
\end{eqnarray}

Real quaternion fields, which preserve time-reversal symmetry was extensively studied in \cite{dNG16}. The authors defined time in real algebra $({\mathcal R})$ and quaternion algebra  $({\mathcal Q})$.
In  ${\mathcal R}$, time reversal corresponds to $u\to \bar u$, while in  ${\mathcal Q}$, $\sigma(u)=Q\cdot \bar u\cdot Q^{-1}$, where
\begin{equation}
Q=\left(\begin{array}{cccccccc}
0&-1&|& & & |& & \\
1&0 & | & & &|& & \\
- &- & - &-& - & -& -&- \\
 & & |&\ddots & & | & & \\
 & & | &  & \ddots & | & & \\
- &- & - &-& - & -& -&- \\
 & & | & & & | &0 &-1 \\
  & & | & & & | &1 &0 \end{array}
  \right)
\end{equation} 
is the symplectic matrix\cite{Chevalley46, Souriau70}. The symplectic group $Sp(n,{\bf R})$ is isomorphic to $GL(2n, {\bf C})$.

For $\epsilon_n$, the unit matrix of degree $n$, let matrix $J$ be
$J=\left(\begin{array}{cc}
0 & \epsilon_n \\
-\epsilon_n & 0\end{array}\right)$ , matrices $\sigma$ that satisfy $^t\sigma J\sigma=J$ are $Sp(n, {\bf C})$ matrices.

The quaternions $\bf H$ has orthogonal bases of the Hlibert space consisting of eigen eiements ${\bf I, i, j ,k}$ which satisfy
\begin{equation}
{\bf  ij=-ji=k,\quad jk=-kj=i, \quad ki=-ik=j}
\end{equation}
In the $SL(2, {\bf C})$ representation
\begin{eqnarray}
{\bf H}&=&a{\bf I}+b{\bf i}+c{\bf j}+d{\bf k}\nonumber\\
&&=\left(\begin{array}{cc}
a+\sqrt{-1}d& -c+\sqrt{-1}b\\
c+\sqrt{-1}b &a-\sqrt{-1}d\end{array}\right),
\end{eqnarray}
where $a,b,c,d\in{\bf R}$ and ${\bf H}^*=a{\bf I}-b{\bf i}-c{\bf j}-d{\bf k}$.

Topology of manifolds on which functions of these numbers are defined turned out to be important to understanding physical phenomena\cite{KR90,Hirzebruch90}. 

\subsection{Supersymmetry in soliton-fermion interactions}
Witten's formulation of supersymmetry was applied to non-linear $\sigma$ model and the effect of instantons in ferromagnetic and anti-ferromagnetic materials were studied. 
A good review on supersymmetry in particle physics and introduction to the super gauge transformation of Wess and Zumino\cite{WZ74} are written in the book of Labelle \cite{Labelle10}.  

Electrons in $2D$ strong magnetic field show Quantum Hall Effects (QHE).  In calculation of conductivity tensors $\sigma_{xy}$ and density of states, supersymmetric analyses was performed. Supersymmetry in disorder and chaos is discussed in \cite{Efetov97}.

In fermionic materials, non-linear interactions sometimes induce solitons which are observed as phonons. Phonons are bosons but do not have supersymmetric partners. In 1 dimensional antiferromagnetic Heisenberg chains, Haldane\cite{Haldane83} showed that spectroscopy of solitons in integer spin system and in half-integer spin system are distinguished. 

In order to study these physical processes, usual methods is to define state space of fermions and bosons using 2nd quantization \cite{Berezin66}. State space of bosons is described by
\begin{equation}
{\mathcal H}_B=\sum_{n=0}^\infty \oplus {\mathcal H}_B^n
\end{equation}
and vectors in the space is described by creation operators $a^*(x_i)$ as
\begin{equation}
\Phi(a^*)=\sum_n \frac{1}{(n !)^{1/2}}\int  K_n(x_1.\cdots ,x_n)a^*(x_1)\cdots a^*(x_n)d^n x,
\end{equation}
where $K_n(x_1,\cdots,x_n)$ are symmetric functions. The annihilation operator $a(x_j)$ is complex conjugate of $a^*(x_i)$.

State space of fermions is described by ${\mathcal H}_F$ is similar to $\Phi(a^*)$, but creation operator and annihilation operator are generators of Grassmann algebra
\begin{equation}
\{ a(x),a(x')\}=\{a(x), a^*(x')\}=\{a^*(x),a'(x)\}=0. \nonumber
\end{equation}
Vectors in the space is expressed as
\begin{equation}
\Phi_{n_1, n_2,\cdots,n_k,\cdots}=\frac{1}{(n_1 ! n_2 ! \cdots)^{1/2}} (a_1^*)^{n_1} (a_2^*)^{n_2}\cdots.\nonumber
\end{equation}

The symmetry between fermions which have a half-integer spin quantum number $s$ and an integer angular momentum number $L$ forming a half-integer total spin quantum number $J$, and bosons which have an integer $s$ and an integer $L$ forming an integer $J$ quantum numbers is the supersymmetry.  Berezin\cite{Berezin66} treated bosons and fermions in Hilbert space not using complex numbers, but generators of $n$ dimensional Grassmann algebra. With each Grassmann algebra, there is closely connected $2n$ dimensional Clifford algebra, whose operators $P_i=\frac{1}{\sqrt{-1}}[\hat x_k-(\partial/\partial x_k)]$, $Q_i=\hat x_k+(\partial/\partial x_k)$ satisfy
\begin{equation}
\{P_i, Q_j\}=0,\quad \{P_i,P_j\}=\{Q_i,Q_j\}=2\delta_{ij}. \nonumber
\end{equation}

The mapping from ${\bf R}^{1,1}\to {\bf H}$ given by 
\begin{eqnarray}
&&[(x_1 e_1+x_2 e_2)=\left(\begin{array}{c}
x_1\\
x_2\end{array}\right)]\nonumber\\
&&\to [x_1{\bf i}+x_2 {\bf j}=\left(\begin{array}{cc}
                                             0 & -x_1+x_2\\
                                             x_1+x_2 & 0\end{array}\right)]
\end{eqnarray} 
is a Clifford algebra.

An aim of this presentation is to show that the super group ${\bf R}^{1,1}$ plays the common role in supersymmetric studies of elementary particles and in solitons.

The structure of this presentation is as follows. In section 2, we discuss supersymmetry in hadron dynamics. In section 3, supersymmetry in low dimensional spin systems is discussed. A bosonic excitation in fermionic systems sometimes appear as solitons. Symmetry in propagation of solitary waves in $2D$ media is discussed in section 4.   In section 5, we present discussion on future research and conclusion.

\section{Supersymmetry in hadron dynamics}
\subsection{Light front holographic QCD in anti de Sitter space}
To construct renormalizable standard model of Quantum Chromo Dynamics (QCD)\cite{Weinberg82}, Srivastava and Brodsky \cite{SB01,SB02} proposed light-front (LF) quantization in the light-cone (LC) gauge, using the Dirac method\cite{Dirac49}.  In this framework and `t Hooft renormalization gauge\cite{tHooft71,tHV72}, they extended the Glashow-Weinberg-Salam model of electroweak interaction without including ghosts that appear in loop calculations.
In LF-form, the vacuum state is trivial up to possible $k^+=k^0+k^z=0$ zero mode, in contrast to the Higgs zero-mode in instant-form.

To include gravitation field in the standard model, Bousso conjectured that local field theory fails and holographic principle, i.e. restricting number of fundamental degrees of freedom in $4D$ space-time is related to the area of surface in $5D$ anti de Sitter space (AdS)  should be utilized\cite{Bousso02}. The gauge/string duality in low energy hadron study is discussed in\cite{BdT06,RT06}. 

AdS space is the maximally symmetric space-timewith negative scalar curvature. $AdS_{d+1}$ space is defined by a surface
\begin{equation}
X_{-1}^2+X_0^2-X_1^2-\cdots-X_d^2=R^2\nonumber
\end{equation}
and the metric of distance
\begin{equation}
ds^2=dX^2+dX_0^2-dX_1^2-\cdots-dX_d^2.\nonumber
\end{equation}
The plane $X_{-1}=X_d$ splits the $AdS_{d+1}$ into two regions defined by light cone coordinates
\begin{equation}
u=\frac{1}{R^2}(X_{-1}-X_d), \quad v=\frac{1}{R^2}(X_{-1}+X_d).\nonumber
\end{equation}

The correspodence between  10-dimensional super gravity and super Yang-Mills equations on the boundary, which is conformal field theory (CFT) was called $AdS/CFT$ correspondence\cite{Maldacena98,SL05}.

de T\'eramond et al.\cite{dTDB14, BdTDE15} extended the LF quantization in the LC gauge based on the method of \cite{dAFF76} to the light front space-time with holographic embedding, by using holographic coordinates
\begin{equation}
z=R^2/(X_{-1}-X_d).\nonumber
\end{equation}

In their formulation, the spectrum of squared mass $M^2$ of mesons and baryons show a similarity. In \cite{dTDB15}, superconformal quantum mechanics of \cite{FR84} applied to the fermionic light-front bound state equations was shown to be dual to the bosonic $AdS_5$. The LF wave function is $\Psi(x_i, {\bf k}_{\perp i},\lambda_i)$ where
\begin{equation}
x_i=k^+/P^+|_i=(k^0+k^z)/(P^0+P^z)|_i\nonumber
\end{equation}
is the LC fraction of a quark, ${\bf k}_{\perp i}$ are the transverse momenta, and $\lambda_i$ are spin projections.
Metric of $AdS$ is
\begin{equation}
ds^2=\frac{R^2}{z^2}(dx^+ dx^- -d{{\bf x}_\perp}^2-dz^2),\nonumber
\end{equation}
where $x^\pm=x^0\pm x^3$ and $z$ is the holographic variable. 

Bousso\cite{Bousso02} and Susskind and Lindesay\cite{SL05} explain utility of holographic principles in AdS space.

The correspondence between the LF holographic QCD (LFHQCD) and supersymmetric quantum mechanism is called $AdS/QCD$ correspondense. Since the longitudinal mode is separated from the transverse modes and the parameter of evolution is taken to be $\tau=x^+=t+z/c$\cite{dTB05, Brodsky06, Brodsky16}. 
Using the LF time evolution operator $P^-=P^0-P^z={d}/{d\tau}$, and $P^+=P^0+P^z$, the LF hamiltonian that operates on $\Psi(x_i, {\bf k}_{\perp i},\lambda_i)$ is given by \cite{Brodsky18, Brodsky19, Sandapen20},
\begin{eqnarray} 
H_{LF}&=&P^+ P^--{\vec P_\perp}^2 \nonumber\\
&=&\frac{{{\bf k}_\perp}^2+m^2}{x(1-x)}+U_{eff}(x,{\bf k}_{\perp }).
\end{eqnarray}

As mentioned by Srivastava and Brodsky\cite{SB01,SB02}, Faddeev-Popov\cite{FP67} or Gupta-Bleuler\cite{Bleuler86} ghost terms are absent in the LCHQCD, and the mass scale of hadrons given by
\begin{equation}
H_{LF}|\Psi(x,{\bf k}_\perp, \lambda)\rangle={\mathcal M}^2|\Psi(x,{\bf k}_\perp, \lambda)\rangle
\nonumber
\end{equation}
are defined by a parameter in $U_{eff}(x, {\bf k}_{\perp })$.

In the LFHQCD approach, Brodsky and de T\'eramond found low energy hadron spectrum of the Regge behavior ${\mathcal M}^2\sim n+L$\cite{dTB05}.  The behavior is not identical to the Regge behavior of open strings ${\mathcal M}^2\sim \frac{n-\alpha(0)}{\alpha'}$\cite{GSW86, GSW87}.  Needs of better understandings of chiral symmetry breaking and confinement as pointed out by t'Hooft\cite{tHooft06} were discussed.
 Waves of relativistic dynamics can be expressed in the instant form, the front form and the point form \cite{ BdTDE15,Dirac49, Brodsky06}. The front form is the wave surface expressed by $u_0-u_3/c=0$. 

When a (1+2)D front form wave function is expressed by quaternions $q\ne 0$, equivalent quaternions that satisfy $q_1 q=q q_2$ have the periodicity in $\tau$ direction by $2 l_{mn}/c$.  When phonons have effective mass due to scattering in media, the wave front in instatnt form changes to the point form.


In LFHQCD, the Lagrangian in $AdS_{d+1}$ space with $AdS$ radius $R$, and a dilaton background $\phi$, yields the effective action 
\begin{eqnarray}
S_{eff}&=&\int d^dx dz \sqrt{g}e^{\phi(x)}g^{N_1N_1'}\cdots g^{N_J N_J'}\nonumber\\
&&\times(g^{MM'}D_M\Phi_{N_1,\cdots N_J^*}D_M'\Phi_{N_1' \cdots N_J'}\nonumber\\
&&\quad -\mu_{eff}^2(z)\Phi_{N_1\cdots N_J}\Phi_{N_1'\cdots N_J'}),
\end{eqnarray}
where  $\sqrt g=(R/z)^{d+1}$ is the conformal scale, and $D_M$ is the covariant derivative which includes the affine connection.

Witten\cite{Witten98} considered $AdS_{d+1}$ space in Kaluza-Klein spacetime and in manifold $M={S}^1\times {S}^{d-1}$. $AdS_{d+1}$ is described as
\[
u v  \sum_{i=1}^d x_i^2=1
\]
and specify spaces with $u,v>0$ as $AdS_{d+1}^+$, and define a group ${\bf Z}$ generated by
\[
u\to \lambda^{-1}u, \quad v\to \lambda v, \quad x_i\to x_i,
\]
with $\lambda$ a fixed real number. The quotient $AdS_{d+1}^+/{\bf Z}$ is defined as $X_1$. Choosing $1\leq  v\leq \lambda$, with $v=1$ and $v=\lambda$ identified.  $X_1$ is topologically ${\bf R}^4\times { S}^1$

At infinity of $X_1$
\[
u\, v \sum_{i=1}^d {x_i}^2=0
\]
and $u,v$ and $x_i$ can be regarded as homogeneous coordinates, and set
\[
\sum_{i=1}^d x_i^2=1.
\]
The boundary of $X_1$ is a copy of $M={ S}^1\times { S}^{d-1}$.
Boundary conformal field theory on ${ S}^1\times { S}^{d-1}$ is to be compared with that of $X_1={ S}^1\times {\bf R}^d$.

In the AdS/CFT correspondence of Maldacena, Eucledian metric
\[
-\tilde X_{-1}^2-\tilde X_0^2+\tilde X_1^2+\cdots +\tilde X_d^2=-R^2
\]
and 
\begin{eqnarray}
U&=&\tilde X_{-1}+\tilde X_d, \nonumber\\
V&=&\tilde X_{-1}-\tilde X_d=\frac{x^2 u}{R^2}+\frac{R^2}{u},\nonumber\\
 x_\alpha&=&\frac{\tilde X_\alpha R}{U}, \quad \alpha=0,1,\cdots, d-1 
\end{eqnarray}
was defined.

\subsection{Tetraquark states or colour singlet quark pair states}
Negative chilarity component of quark in a meson which yield negative chirality leptons, could produce negative chilalrity component of quarks in baryons. In LFHQCD, a fermion operator $R_\lambda^\dagger$ that connects $q\bar q$ mesons to $qqq$ baryons can create tetra quark $qq\bar q \bar q$ states\cite{BdTDL16}.

When $qq\bar q \bar q$ states dominate over $q\bar q q\bar q$ states is an interesting problem for the study of supersymmetry. Brodsky et al\cite{BdTDL16} emphasized that the supersymmetric features of hadron physics derived from superconformal quantum mechanics refers to the symmetry properties of bound state wave functions of hadrons and not to quantum fields. Similar argument based on ontological quantum mechanics was proposed also by `t Hooft\cite{tHooft20}.

Recently the CMS collaboration at LHC announced detection of $\Upsilon(1S)\Upsilon(1S)\to\mu^+\mu^-\mu^+\mu^-$ in proton-proton collisions at $\sqrt{s}=13$ TeV \cite{CMS20}.

$\Upsilon(1S), \Upsilon(2S), \Upsilon(3S)$ are $b\bar b$ bosons of $J^{PC}=1^{--}$ and have mass
9.460 GeV, 10,023 GeV and 10.355 GeV, respectively. There is also $\Upsilon(1D)$ of $J^{PC}=2^{--}$, but its decay to $\mu^+\mu^-$ is not detected.

The 4 $\mu$ decay modes seems to be dominated via $\Upsilon(1S)\Upsilon(1S)$ resonances and sum of  single parton scattering (SPS) and double parton scatterng (DPS) fits the experimental data quite well, and tetraquark contribution seems to be small, although there are uncertainty in the quark spin polariztion of $\Upsilon(1S)$. 

When the $b$ quark is replaced by $c$ quark, $c\bar c$ makes $J/\psi$ of $J^{PC}=1^{--}$, and $J/\psi(1S), J/\psi(2S)$ have masses 3.097 GeV and 3.686 GeV, respectively.  $c\bar c$ contribution to the proton electromagnetic form factor were calculated using lattice data and $s\bar s$ contribution was discussed\cite{SLABdTDDLY20}.  Similar analysis will become possible if $b\bar b$ amplitudes are similar to those of $c\bar c$.

Concerning $J/\psi J/\psi$, $\Upsilon \Upsilon$ and $J/\psi\Upsilon$ productions, there is a proposal of single and double quarkonium productions in the colour-evapolation model\cite{LSYZN20}, which is an extension of a model for $J/\psi J /\psi$ production\cite{LS13}. The colour evapolation model (CSM) assumes direct production of vector quarkonium from gluons.

In the case of production of $J/\psi J/\psi$, the cross section per quarkonium transvers momentum
$d\sigma/d P_T$ in perturbative QCD was found to be larger than that of $J/\psi+\eta_c+g$ including QCD coupling conststants $\alpha_s(m_q)^5$. Experiments at LHC indicate that NLO $\alpha(m_q)^4$ and NNLO $\alpha_s(m_q)^5$ were found to be larger than the leading order (LO)\cite{Landesberg19}.

The calculated cross sectiion of $\Upsilon\Upsilon$ production in CSM turned out to be smaller than that obtained by the sum of SPS and DPS evaluated by the CMS collaboration. There remain problems on tetra quark models and quark pair models.

\subsection{Noncommutative geometry approach}
Connes and Lott\cite{CL90} proposed a quantum field theory of Yang-Mills type equations\cite{YM54}, by adopting a gauge fixing excluding ghost fields, based on noncommutative geometry\cite{Connes90,Connes94}.

We define the complex projective space $P_m$ by taking $C_{m+1}-0$, where $C_{m+1}$ is $(m+1)$-tuples $(z_0,z_1,\cdots,z_m)$ and $0$ is the point $(0,\cdots,0)$\cite{Chern67}. A natural projection 
\begin{equation}
\psi: C_{m+1}-0\to P_m
\end{equation}
and inverse image of each point is $C^*=C_1-0$. In $\psi^{-1}(U_i)$ where $U_i$ is an open neighbourhood of $z^i$ we use coordinates  $_i\zeta^h=z^h/z^i$, $(0\leq h\leq m, h\ne i)$, and $z^i$.  
 To a point $p\in P_m$, $\psi^{-1}(p)$ are called its homogeneous coordinates. 
 
 The equation
 \begin{equation}
 \sum_{k=0}^{m} z^k \bar z^k=1
 \end{equation}
 defines a $S^{2m+1}$ sphere. The mapping $\psi : S^{2m+1}\to P_m$, whose inverse image of each point is a circle is called Hopf fibering of $S^{2m+1}$.
 
 Let $\Gamma$ be the discontinuous group generated by $2m$ translations of $C_m$. Then $C_m/\Gamma$
 is called the complex torus.
 
 Let $\Delta$ be the discontinuous group generated by $z^k\to 2z^k$, $(1\leq k\leq m)$. The quotient manifold 
 $(C_m-0)/\Delta$ is called the Hopf manifold. It is homeomorphic to $S^1\times S^{2m-1}$.

 The Dirac matrices, which are familiar to physicists represent Clifford algebra ${\mathcal A}_{3,1}$ which is a subalgebra of ${\mathcal A}_{3,2}\sim M_4({\bf C})$.
The algebra ${\mathcal A}_{3,1}$ is isomorphic to $M_2({\bf H})$, where $\bf H$ is the quaternion field\cite{Garling11}.
Spinor structures and Clifford modules were studied in \cite{ABS64} and conformal transformations of Dirac spinors were used in \cite{WZ74}, and relations to complex Grassmannian are explained in \cite{Berezin66}.  

In an article of Connes et al\cite{Connes17}, an extension of noncommutative geometry formalism to unify gravity and the supersymmetric standard model\cite{CC96} in Einstein-Hilbert space was discussed. 
He defined spectral triple $({\mathcal A}, {\mathcal H}, D)$, where ${\mathcal A}$ is the algebra generated by ${\bf R},{\bf C}$ and ${\bf H}$, $\mathcal H$ is the Hilbert space and $D$ is a self-adjoint operator.

By a proper choice of foliation $F$ in the space $V={\bf R}^n$ of noncommutative algebra,  longitudinal components could be eliminated from effective cohomology, and extract transverse components in the Hilbert space, and ghosts were not required for construction of consistent gauge theory \cite{Connes94}. 

Supersymmetric field theory on noncommutative geometry is based on the supergroup ${\bf R}^{1,1}$. 
The Hopf algebra $H$ of smooth functions on the ${\bf R}^{1,1}$ is given as
\begin{equation}
H=C^\infty({\bf R}^{1,1})=C^\infty({\bf R})\otimes \wedge({\bf R})
\end{equation}\nonumber
where $\wedge({\bf R})$ is the exterior algebra of one-dimensional vector space. Every element of $H$ can be expressed as $f+g\theta_1$, where $f,g\in C^\infty({\bf R})$ and $\theta_1^2=0$. Thus supersymmetry becomes manifest in $H$. In terms of Grassmann variables $\theta=(\theta_1,\theta_2)$, $F(\theta_1)=c_0+c_1\theta_1$ has the similar structure as $H$. 

Jaffe et al \cite{JLL87,JLW87,JLO88} extended the ${\mathcal N}=2$ supersymmetric Wess-Zumino quantum fields adopted by Witten\cite{Witten82c}, using an extension of Connes' triple as spectral quadruple $({\mathcal A},{\mathcal H},\Gamma, Q)$ where $\Gamma$ is ${\bf Z}_2$ grading of $\mathcal A$ and $Q$ is the Dirac operator,
\begin{equation}
Q=\left(\begin{array}{cc}
0& Q_-\\
Q_+& 0\end{array}\right),
\end{equation}
where $Q_-$ operates on ${\mathcal H}_-\to{\mathcal H}_+$ and  $Q_+$ operates reversally.

The Hamiltonian $H=Q^2$ is
\begin{equation}
H=\left(\begin{array}{cc}
Q_+^* Q_+& 0\\
0 & Q_-^*Q_-\end{array}\right),
\end{equation}
and considered the heat kernel $e^{-\beta H}$ $(\beta>0)$, state vectors $a(t)=e^{-tH}a e^{tH}$ and cochains ${\mathcal C(A)}$ consisting of $(f_0,f_1,f_2,\cdots)$. 
Grading of $\Gamma$ allows decompositions of ${\mathcal C(A)}$
\begin{equation}
{\mathcal C(A)}={\mathcal C}^e({\mathcal A})\oplus {\mathcal C}^o({\mathcal A}),
\end{equation}
where $(f_0,f_2,f_4,\cdots)\in{\mathcal C(A)}^e$, $(f_1,f_3,f_5,\cdots)\in{\mathcal C(A)}^o$,
and 
\begin{equation}
{\mathcal C(A)}^e=\frac{1+\Gamma}{2}{\mathcal C(A)}^e+\frac{1-\Gamma}{2}{\mathcal C(A)}^e={\mathcal C(A)}^e_+ +{\mathcal C(A)}^e_-\nonumber.
\end{equation}

They showed that the coboundary operator $\partial$ exists that show the cohomology sequence
\begin{equation}
\cdots\to {\mathcal C}_+^e({\mathcal A})\stackrel{\partial}{\to} {\mathcal C}_+^0({\mathcal A})\stackrel{\partial}{\to} {\mathcal C}_+^e(.{\mathcal A})\to\cdots
\end{equation}

In the case of ${\mathcal N}=2$ Wess-Zumino supersymmetric model, one can choose
\begin{eqnarray}
Q_1&=&\sqrt{-1}\bar\psi_1\partial -\sqrt{-1}\bar\psi_2(\partial V)^*, \nonumber\\
Q_2&=&\sqrt{-1}\psi_2\bar\partial+\sqrt{-1}\psi_1\partial V,
\end{eqnarray}
and 
\begin{equation}
H=Q^2=-\partial\bar\partial-\bar\psi_1\psi_1\partial^2V-\bar\psi_2 \psi_2(\partial^2 V)^*+|\partial V|^2,
\end{equation}
is hermitian due to $\bar\psi_1\psi_1=(\bar\psi_2\psi_2)^*.$

One can consider spin 1/2 fermion $SU(2)$ gauge theory with Yukawa coupling, but the pion remains massless.

\subsection{Supersymmetry and the BRST charge}
In topological quantum field theory, Witten defined the charge of Becchi-Rouet-Stora-Tyutin (BRST) transformations\cite{BRS75, Tyutin75} which contain Faddeev-Popov ghosts as follows\cite{Witten88}. 

One begins with gauge fields $A_i^a(x)$ on a three manifold $Y$. Here $i=1,2,3$ labels the components of a tangent vector to a manifold $M$, $a$ runs over the generators of a gauge group $G$, $x$ labels a point in $Y$, which is endowed with a metric tensor $g_{ij}$. The variation $\delta A_i^a(x)$ can be regarded as a differential operator on differential forn $\omega$ on the space $\mathcal A$: $\omega\to \delta A^a_i(x)\wedge\omega$.
The exterior derivative and its adjoint on $\mathcal A$ are
\begin{equation}
d=\int d^3 x\psi_i^a(x)\frac{\delta}{\delta A_i^a(x)}, \quad d^*=-\int d^3 x\chi_i^a(x)\frac{\delta}{\delta A_i^a(x)},
\end{equation}
where $\chi_i^a(x)$ is a vector dual to $\psi_i^a(x)$. With a real parameter $t$, and Chern-Simons functional
\begin{equation}
W=\frac{1}{2}\int_Y Tr(A\wedge dA+\frac{2}{3}A\wedge A\wedge A),\nonumber
\end{equation}
one defines
$d_t=e^{-tW}d e^{tW}$, $d_t^*=e^{tW}d^* e^{-tW}$ that satisfy
\begin{equation}
d_t^2=0, \quad {d_t^*}^2=0,\quad d_t d_t^*+d_t^* d_t=2H.\nonumber
\end{equation}
The group of $Y$ is graded and additive quantum number $U$ is -1 for $\chi$ and +1 for $\psi$. 

In order to obtain a relativistic version, one works on space $Y$ and time $R^1$, and in the space $Y\times R^1$ name the supersymmetric generator $d_t$ as $Q$.
One tries the Lagrangian
\begin{equation}
\mathcal L=\int_M dx Tr [\frac{1}{4}F_{\alpha\beta}F^{\alpha\beta}-i\eta D_\alpha\psi^\alpha+i(D_\alpha\psi_\beta)\chi^{\alpha\beta}].\nonumber
\end{equation}
The supersymmetry transformation law is
\begin{eqnarray}
&&\delta A_\alpha=i\epsilon\psi^\alpha,\quad \delta\eta=0,\quad \delta\psi^\alpha=0,\nonumber\\
&&\delta\chi_{\alpha\beta}=\epsilon(F_{\alpha\beta}+\frac{1}{2}\epsilon_{\alpha\beta\gamma\delta}F^{\gamma\delta}).\nonumber
\end{eqnarray}
The propagating modes of $A_\alpha$ have helicities $(1,-1)$, and those of $(\eta,\psi,\chi)$  have helicities $(1,-1,0,0)$.

The Lagrangian invariant under the fermionic symmetry
\begin{eqnarray}
&&\delta A_\alpha=i\epsilon\psi_\alpha,\quad \delta\phi=0,\quad \delta\lambda=2i\epsilon\eta,\quad \delta\eta=\frac{1}{2}\epsilon[\phi,\lambda],\nonumber\\
&&\delta\psi_\alpha=-\epsilon D_\alpha\phi,\quad \delta\chi_{\alpha\beta}=\epsilon(F_{\alpha\beta}+\frac{1}{2}\epsilon_{\alpha\beta\gamma\delta}F^{\gamma\delta}),
\end{eqnarray}
where $\lambda$, $\phi$ are new scalar fields, becomes
\begin{eqnarray}
&&{\mathcal L}_0=\int_M d^4 x Tr[\frac{1}{4}F_{\alpha\beta}^{\alpha\beta}+\frac{1}{2}\phi D_\alpha D^\alpha\lambda\nonumber\\
&&+iD_\alpha\psi_\beta\cdot \chi^{\alpha\beta}-\frac{i}{8}\phi[\chi_{\alpha\beta},\chi^{\alpha\beta}]-\frac{i}{2}\lambda[\psi_\alpha,\psi^\alpha]].
\end{eqnarray}

For any functional $\mathcal O$, its variation under the fermionic symmetry becomes 
\begin{equation}
\delta{\mathcal O}=-i\epsilon\cdot\{Q,{\mathcal O}\}.\nonumber
\end{equation}
For ${\mathcal O}=\frac{1}{4}Tr([\phi,\lambda]\eta)$, there is a possibility to add a Lagrangian
\begin{equation}
{\mathcal L}_1=s\int d^4 x \{Q,\mathcal O\}=s\int d^4 x Tr[\frac{i}{2}\phi[\eta,\eta]+\frac{1}{8}[\phi,\lambda]^2]\nonumber
\end{equation}
with $s$ an arbitrary parameter.

By choosing $s=-1$, the relativistic Lagrangian becomes
\begin{eqnarray}
&&{\mathcal L}=\int_M d^4 x{\sqrt g}Tr[\frac{1}{4}F_{\alpha\beta}F^{\alpha\beta}+\frac{1}{2}\phi D_\alpha D^\alpha \lambda \nonumber\\
&&-i\eta D_\alpha\psi^\alpha+i D_\alpha\psi_\beta\cdot\chi^{\alpha\beta}-\frac{i}{8}\phi[\chi_{\alpha\beta},\chi^{\alpha\beta}]\nonumber\\
&&-\frac{i}{2}\lambda[\psi_\alpha,\psi^\alpha]-\frac{i}{2}\phi[\eta,\eta]-\frac{1}{8}[\phi,\lambda]^2].
\end{eqnarray} 

One defines a tensor $T_{\alpha\beta}$ such that under an infinitesimal change of metric $g^{\alpha\beta}\to g^{\alpha\beta}+\delta g^{\alpha\beta}$, the change of the action becomes
\begin{equation}
\delta{\mathcal L}=\frac{1}{2}\int_M \sqrt g\delta g^{\alpha\beta}T_{\alpha\beta}.
\nonumber
\end{equation}
$T_{\alpha\beta}$ can be expressed in the form of BRST commutator
\begin{equation}
T_{\alpha\beta}=\{Q,\lambda_{\alpha\beta}\}\nonumber
\end{equation}
with
\begin{eqnarray}
&&\lambda_{\alpha\beta}=\frac{1}{2}Tr(F_{\alpha\sigma}\chi_\beta^\sigma+F_{\beta\sigma}\chi_\alpha^\sigma-\frac{1}{2}g_{\alpha\beta}F_{\sigma\tau}\chi^{\sigma\tau}\nonumber\\
&&+\frac{1}{2}Tr(\psi_\alpha D_\beta\lambda+\psi_\beta D_\alpha\lambda-g_{\alpha\beta}\psi_\sigma D^\sigma\lambda)\nonumber\\
&&+\frac{1}{4}g_{\alpha\beta}Tr(\eta [\phi,\lambda]).
\end{eqnarray}
One can show that $D_\alpha T^{\alpha\beta}=0$ \cite{Witten88}.

In contrast to the $Q_{BRST}\sim c\partial cb+c(\partial X)^2$, where $c,b$ are conformal ghosts and $X$ is the matter field of Kugo and Ojima\cite{KO78,KO79}, the supercharge $Q$ is defined on twisted manifods $M=Y\times R^1$, and ghosts are absent in the supersymmetric theory.

We investigated the Kugo-Ojima parameter $u(0)=c$ in lattice Landau gauge QCD, and observed a small deviation from $c=1$\cite{NF99, NF04}. In lattice simulations, it is necessary to extrapolate finite size simulation data to continuum limit, and it is difficult to check whether nature satisfies the BRST symmetry\cite{FN06, SF20}. 

Schaden and Zwanziger\cite{SZ15} studied BRST cohomology by choosing physical subspace $\mathcal V$ to be the physical sector of the Gribov-Zwanziger (GZ) theory\cite{Zwanziger93}. The GZ vacuum breaks BRST symmetry and
the Lagrangian is written as
\begin{equation}
{\mathcal L}={\mathcal L}^{YM}+{\mathcal L}^{ghosts}={\mathcal L}^{YM}+s\Psi\nonumber
\end{equation}
The expectation value  $\langle s\Psi\rangle$ becomes zero due to cancellation of fermion ghosts and boson ghosts.

Watson and Alkofer\cite{WA18} verified the Kugo-Ojima confinement criterion with an assumption that the physical subspace is $\mathcal V=Ker( Q_{BRST})$ and physical states are given by the cohomology ${\mathcal H}(Q_{BRST},{\mathcal V})=Ker( Q_{BRST})/Im( Q_{BRST})$. They showed the leading infrared powers of the gluon $\delta_0$ and the ghost propagator $\epsilon_0$ are related, and the enhancement of the ghost propagator and suppression of the gluon propagator in infrared region as compared to tree level calculations can be understood. 

There is an attempt to construct BRST complex that gives Gupta-Bleuler space of physical states using Cohomological reductions\cite{HC20}. Ghosts can be regarded as supersymmetric partners of fermions that were required to keep 4D space inside the Gupta-Bleuler space.
Theories based on quantum K-theory\cite{JLO88, Connes94} in which ghosts are absent may be more favorable to understand symmetries of nature. 

Experimental check of the standard model, or searches of theories beyond the standard model was performed at LHC, Belle and other laboratories through two photon emission measurements. At LHC gluon fusions $gg\to X\to \gamma\gamma$ where $X$ scalar bosons including 125GeV Higgs boson.  The standard model of the Higgs spectrum is based on $SO(5)/SO(4)$ coset. 

The group $SO(n)$ is a kernel of determinant map $O(n)\to \{\pm 1\}$. 
$SO(3)$ is homeomorphic to projection space ${\bf RP}^3$\cite{Hatcher01}. $SO(4)$ is homeomorphic to $S^3\times SO(3)$. Identifying ${\bf R}^4$ with the quaternions $\bf H$ and $S^3$ as the unit of quaternions,
one obtains the homeomorphism $S^3\times SO(3)\approx SO(4)$\cite{Hatcher01}. The homomorphism $\psi: S^3\times S^3\to SO(4)$ that sends a pair of unit quaternions 
\begin{equation}
(u,v)=(u_0{\bf I}+u_1{\bf i}+u_2{\bf j}+u_3{\bf k},v_0{\bf I}+v_1{\bf i}+v_2{\bf j}+v_3{\bf k})\nonumber
\end{equation}
to the isometry $w\to u w v^{-1}$ of ${\bf H}$ is surjective with the kernel ${\bf Z}_2=\{\pm(1,1)\}$.

Using octonions $\bf O$, one can construct a homeomorphism $SO(8)\approx S^7\times SO(7)$, but in other $n$, $SO(n)$ is only a twisted product of $S(n-1)\times S^{n-1}$\cite{Hatcher01}.
As an example, $SO(5)$ is homeomorphic to a twisted product of $SO(4)$ and $S^4$,  $SO(5)/SO(4)$ is homeomorphic to $S^4\times {\bf Z}_2$. 

When Kaluza-Klein 5 dimensional vector gauge theory works, an extension of the Higgs model based on $SO(5)/SO(4)$ which includes the WZW term is possible\cite{KMPZ16,ACCDLCS17}.  The model is based on spontaneous $CP$ symmetry breaking mechanism of Peccei and Quinn (PQ)\cite{PQ77}. 

The PQ model is a model of fermions which couple to non-abelian gauge fields, whose effective action is given by
\begin{equation}
S_{eff}^q=\int d^4x( {\mathcal L}+\sqrt{-1}\theta q),
\end{equation}
where $q=(g^2/32\pi^2)\int d^4 x F_{\mu\nu}^a \tilde F^{a\mu\nu}$.
 
The rotation of the fermion field by $exp(\sqrt{-1}\gamma_5 \eta)$ induces a change in the effective action
\begin{equation}
\delta S_{eff}^q =-\sqrt{-1}\int d^4 x (\partial^\mu j_\mu^5)\eta=-2\sqrt{-1}q \eta
\end{equation}
and the net effect is the rotation $\theta\to \theta-2\eta$. It induces spontaneous symmetry breaking which creates Higgs bosons. 

In topological classification of $2n$ free fermions systems \cite{FK10}, 4 Majorana fermion system $\hat c_1 \hat c_2 \hat c_3 \hat c_4$ with the $so(3)$ symmetry, and 8 Majorana fermion system $\hat c_1,\cdots, \hat c_8$ with the $so(8)$ symmetry were studied. In the $so(8)$ space, a subspace ${\bf 8}_+$ and a subspace ${\bf 8}_-$ which are invariant under $so(7)$ were chosen and the triality symmetry of octonions were taken into account. 

The Hopf theorem says that there exists a continuous mapping of $S^{n-1}\times S^{n-1}\to S^{n-1}$ where $n$ is a power of 2. When $n=4$, the mapping $S^3\times S^3\to S^3$ can be expressed by transformations of three unit vectors in $S^3$ represented by quaternion algebra. When $n=8$, $S^7\times S^7\to S^7$ is represented by octonions, and this $n=8$ is largest due to the Bott periodicity\cite{BT82,Hatcher01}.  
 
The PQ model predicts Goldstone boson \cite{Kibble67} and axion like particles\cite{JJS13}. Axion-like particles are (pseudo) scalar particles $\phi$ of mass $m_\phi$ interacting with the standard model electro-weak gauge field-tensor $B_{\mu\nu}$ and $G_{\mu\nu}$ whose interaction Lagrangian is
\begin{equation}
{\mathcal L}_\phi\supset -\frac{1}{4}g_{\phi BB} \phi B_{\mu\nu}\bar B_{\mu\nu}-\frac{1}{4}g_{\phi gg} \phi G_{\mu\nu}\bar Higg G_{\mu\nu},\nonumber
\end{equation}
where $B_{\mu\nu}$ is electro weak gauge field-tensor and  $G_{\mu\nu}$ is the Goldstone boson field-tensor.

\subsection{Extensions of Higgs boson model and supersymmetry}
In minimal supersymmetric standard model (MSSM)\cite{Labelle10}, Higgs field consists of $SU(2)_L$ doublet
\[
H_u=\left(\begin{array}{c} {H_u}^+\\
                             {H_u}^0\end{array}\right)\geq \left(\begin{array}{c} 0\\
                                                                                            \nu_u\end{array}\right)
\]
and 
\[
H_d=\left(\begin{array}{c} {H_d}^0\\
                             {H_d}^-\end{array}\right)\geq \left(\begin{array}{c} \nu_d\\
                                                                                            0\end{array}\right)
\]
where $H^+, H^0$ are complex scalar fields.  The Higgs bosons in the MSSM consists of three massless
and five massive ($h^0, H^0, H^{\pm}, A^0$), and $M_h^0\sim 125 $GeV.

In \cite{BLS19} the $\tan\beta$ parameter in the MSSM is defined as
\[
\tan\beta=\frac{\nu_u}{\nu_d}
\]
The partner of the standard model fermion are expected to exist beyond TeV scale and $\tan\beta<5$ is ruled out.

 In the flamework of search of beyond the standard model, there is the two Higgs doublet model (THDM)\cite{BCGPPRRV18,BH18} as an effective field theory below the SUSY scale. In this model, renormalization on $\tan \beta$ was reconsidered by removing sleptons and squarks.

In the standard model, $CP$ violation occurs through Kobayashi-Maskawa $SU(3)$ matrix phases\cite{KM73} in $c$ quark and $s$ quark dynamics $D^0\to K^+ K^-$\cite{Miller19,LHCb19} and $D^0\to \pi^+\pi^-$\cite{LHCb12}.

Authors of \cite{BCGPPRRV18} studied $CP$ violation which can be detected in 500 GeV $e^+ e^-$ collision in International Linear Collider (ILC) and 380 GeV $e^+e^-$ collision in Compact Linear Collider (CLIC).

In the mathematical model of strings with Dirichret boundary condition (D-Brane), one can assume the flipped $SU(5)\times U(1)$model for hadron interactions\cite{DBLLLMN18} and estimate effects of supersymmetric partner of matters to the mass of Higgs bosons etc.

\subsection{Dark Matter and supersymmetry}
Duan et al.\cite{DHRWY18} proposed a probe of bino-wino coannihilation dark matter below the neutrino floor at the LHC. 
Winos are SUSY partner of $W^{\pm}$ of the MSSM theory, and the proton-proton collision in LHC that creates winolike $\chi_2^0$
\[
p p \to {\chi_2}^0(\to l^+ l^- {\chi_1}^0){\chi_1}^{\pm}+{\rm jets}
\]
was considered. Here $\chi_1^0$ is the lightest neutralino is a weakly interacting massve particle (WIMP) dark matter (DM) which is a linear combination of gauge eigenstates $(\bar B, \bar W, \bar H_d^0, \bar H_u^0)$, i.e.
\[
{\chi_1}^0=N_{11}\bar B+N_{12}\bar W+N_{13}{\bar H_d}^0 +N_{14}{\bar H_u}^0.
\]
When $N_{11}^2>max\{ N_{12}^2, N_{13}^2+N_{14}^2\}$, ${\chi_1}^0$ is called binolike. 
 
Possibility of detecting DM relic density is discussed.

Relations between DM and Big Bang Nucleosynthesis (BBN) was studied by Hamdan and Unwin\cite{HU18}. In the BBN, DM decomposition is assumed to occur during radiation domination. In presence of additional dimension to 4 dimensional space-time, DM can decouple when the matter-like universe exists.

\subsection{Non-local deformation of a supersymmetric field theory}
Heisenberg uncertainty problem is related to the length scale of physics and it is an important ingredient for $\mathcal N=1$ supersymmetric field theory and conformal field theory \cite{ZFSBGZMRI17}. 

Materials that are considered are Zeeman driven Lifshitz transition, which is a phase transition caused by changes of Fermi surface in a heavy fermion metal \cite{HV11}, and  theoretical bases were taken from \cite{BA12}.

\subsection{New colored scalar bosons}
In the search of beyond the standard model, Gabrielli et al. \cite{GKMRSV16} proposed a colored scalar boson of mass 750GeV, as a cause of bumps of two-photon final state observed at LHC. The LHC Run 1 and Run 2 data at 7TeV and 8 TeV, the cross section was $4.5\pm 1.9$ fb, and the data at 13TeV was $10.6\pm 2.9$ fb.
Their expectations is $gg\to S\to\gamma\gamma$ with triangle diagrams of $S\to g g$ and $S\to\gamma\gamma$. 

Triangle diagrams appear in the theory of anomaly\cite{Fujikawa79}. 

Higgs boson decays to dibosons including $\gamma\gamma$ was measured at LHC\cite{ATLAS13,ATLAS14} and analyzed as an example in \cite{Furui15}.

 \subsection{Superconformal field theory and M-Theory}
 In the string theorry and M-theory, the flux compactifiation in background field (Gravitational field) becomes necessary. Lazaroiu et al \cite{LBC16} studied supersymmetric flux compactification of supergravity theories, by defining constrained generalized Killing spinors. A Killing vector in a Riemann space with metrics $g_{\lambda\mu}$ is written as $X=\xi^\lambda \frac{\partial}{\partial x^\lambda}$ where $\xi$ satisfies
 \begin{eqnarray}
&& \xi^\alpha\frac{\partial g_{\lambda\mu}}{\partial x^\alpha}+g_{\lambda\alpha}\frac{\partial \xi^\alpha}{\partial x^\mu}+g_{\alpha\mu}\frac{\partial \xi^\alpha}{\partial x^\lambda}\nonumber\\
&&\nabla_\lambda \xi_\mu+\nabla_\mu \xi_\lambda=0.
 \end{eqnarray}
 
 Killing spinor $\psi$ satisfies
 \[
 \nabla_X\psi=\lambda X\cdot\psi,
 \]
where $\cdot$ is the Clifford multiplication and considered Kaehler-Atiyah bundle. 

Kaehler metric in complex projective spaces are hermitian metric $\omega=\sqrt{-1}\sum_{\alpha,\beta}g_{\alpha\beta} dz^\alpha\wedge d\bar z^\beta$ that satisfies $d\omega=0$.  
 Lazaroiu et al.\cite{LBC16} referred Gauntlett et al.\cite{GMW04} for the Kaehler-Atiyah metric. Gaunett et al  discussed superstring with intrinsic torsion. The are two types of string theories: 1) open string theory and 2) closed string theory. Gravitational effects in string theories can be studied in \cite{GSW86, GSW87}. For an arbitrary manifold, one can not necessarily define a Kaehler metric\cite{Kodaira92}. Gauntlett et al. \cite{GMW04} considered complex structures on the $K3$ manifold also. Their manifold allows intrinsic torsion on non-Kaehler manifold, like Hopf manifold\cite{Kodaira92}.

It is well known that superconductivity is well described by Landau-Ginzburg equation\cite{Kittel66}. Witten used the Landau-Ginzburg description in \\
${\mathcal N}=2$ superconformal model\cite{Witten93, RSV88}, and Davenport and Melnikov studied $(2,2)$ compactified supersymmetry theory with central charge\cite{DSB04} $c=\frac{\mathcal{N(N}+1)}{2}<6$.

Kiyoshige and Nishinaka derived formula of three-point functions of 4-dimensional ($4D$)  ${\mathcal N}=2$ superconformal field theories\cite{KN19} using the Calabi-Yau Metric\cite{GSW87}.  ${\mathcal N}=2$ super-conformal field theories in $4D$ was proposed by Seiberg and Witten \cite{SW94} for introducing electric-magnetic duality, monopole condensate and confinement in Yang-Mills theory, and extended to $SU(3)$ supersymmetric gauge theory by Argyres and Douglas \cite{AD95}. Topological structures of closed strings were studied.  

The conformal field theory (CFT) is a starting point for describing infrared (IR) property of asymptotically free field theory in which the world sheet descibed by $2\times 2$ symmetric tensor 
$h_{\alpha\beta}$ is parametrized as $\eta_{\alpha\beta} e^\phi$, where $e^\phi$ is an unknown conformal 
factor \cite{GSW86}.  A ${\mathcal N}=2$ super-conformal field theory (SCFT) in $4D$ was proposed in \cite{APSW96}.

A $4D$ SCFT on $S^3\times {\bf R}$ for ${\mathcal N}=1,2,4$ was studied by Kinney et al. \cite{KMMR07} and in the case of ${\mathcal N}=4$, AdS/CFT correspondence of entropy of Bogomolnyi-Prasad-Sommerfed (BPS) black hole in strong coupling $AdS_5\times S^5$ frameworks. 

Gadde et al.\cite{GRRY11} showed that the superconformal index (the partition function on $S^3\times S^1$) of a certain class of $4D$ supersymmetric field theories is exactly equal to a partition function of $q-$ deformed nonsupersymmetric $2D$ Yang-Mills theory. 

Beem et al \cite{BLLPRR15} classified ${\mathcal N}=2$ superconformal field theory using Schur operators\cite{LBC16}, that operates on quaternions

Using BRST cohomology, they conjectured that $4D$ CFT with extended supersymmetry and $2D$ chiral algebras have a correspondence. 

\subsection{Electron-Boson quantum manybody systems}
In the review article of Caliceti et al.\cite{CMHRSJ07}, problem of getting physical predictions from slowly convergent series was discussed.
Mera et al.\cite{MPN16} applied hypergeometric resummation technique developed by themselves\cite{MPN15} and showed that in the calclation of self-consistent Dyson equtions of electron-boson quantum many-body system, hypergeometric resummation is better than Pad\'e approximant method.

Authors claim that their method is useful in calculation of physical observables in a non-equilibrium steady state electron-boson quantum many-body systems. Dynamics of electrons interacting with phonons in the presence of applied bias voltage is also discussed.

 \subsection{The thermal tensor network}
Dong et al\cite{DCLL17} analyzed thermal tensor networks of a Ising chain, whose hamiltonian is
\[
H=-\sum_i^L (J {S_i}^\alpha S_{i+1}^\alpha+B {S_i}^z).
\]
They used the nonlocal fermionization transformation:
\begin{eqnarray}
c_i&=&exp[i\pi \sum_{j=1}^{i-1}(S_j^z+\frac{1}{2})]{S_i}^-, \nonumber\\
c_i^\dagger&=&exp[-i \pi \sum_{j=1}^{i-1}(S_j^z+\frac{1}{2})]S_i^+, \nonumber\\
c_i^\dagger c_i&=&S_i^z+\frac{1}{2},\nonumber
\end{eqnarray}
where $S_i^+=S_i^x+\sqrt{-1}S_i^y, \quad S_i^-=S_x-\sqrt{-1}S_i^y$.

The Hamiltonian is transformed to
\[
H=\frac{B L}{2}-B\sum_{i=1}^L c_i^\dagger c_i-\frac{J}{4}\sum_{i=1}^L(c_i^\dagger-c_i)(c_{i+1}^\dagger+c_{i+1}).
\] 
Linearized tensor renormalization group (LTRG) for bilayers was applied to a fermionc extended Hubbard model (EHM).

\section{Supersymmetry in low dimensional spin system}
In 1983, Haldane \cite{Haldane83} showed in one-dimensional non-linear sigma model, difference of half-integer spin states and integer spin states, which have different instability induces solitons.  Couplings of nonlinear zero-point fluctuation (instantons) were discussed. He studied stabilities of half-integer spin fermionic systems and integer spin fermionic systems. 

In the same year, Efetov \cite{Efetov83} studied non-linear supermatrix sigma model, using supervectors and super matrices, which are based on Berezinian (Grassmann algebra for fermions)\cite{Berezin66}.

Altland and Zirnbauer \cite{AZ97} studied mesoscopic junctions of normal-conducting metals and super-conducting metals, using the Efetov's method. They showed that the system is classified by the symmetry operations of time reversal and rotation of electron's spin, and the 4 symmetry classes of Hamiltonian corresponding to time-reversal symmetry (TRS), particle-hole symmetry (PHS) and chiral symmetry (SLS) which is a product space of TRS and PHS, corresponding to Cartan's symmetric spaces of type $C, CI, D$ and $DIII$ which are called Bogolyubov-de Gennes symmetry classes.

Classification of topological insulators and superconductors in three spacial dimension was studied by Schnyder et al.\cite{SRFL08}. 

 In Witten's model, triangle diagrams appear in the argument of symmetry-protected topological (SPT) bosonic phases through Chern-Simons theory \cite{Witten15}.  

Symmetry protected topological phases in two dimensional fermionic system was studied by Gu and Wen\cite{GW14} and gapless modes at boundaries were studied by Zhang et al.\cite{ZWYQG20}.  They classified phases of 2D fermionic symmetry protected topological phase (FSPT) using the technique of induced representations applied to quantum mechanics\cite{Mackey68}.

For spinless fermions, the autors consider cyclic groups $C_{2m-1}$ and $C_{2m}$ which has ${\bf Z}_{2m-1}$ and ${\bf Z}_{2m}$ symmetries, respectively.
 Semidirect products of rotation $C_n$ and reflection symmetry ${{\bf Z}_2}^M$ which are diheadral group $D_n=C_n\ltimes {{\bf Z}_2}^M$. $D_{2m-1}=D_{2m}={\bf Z}_2$.
  
The classification of $C_{2m-1}$ for spin $1/2$ fermions is same as spinless fermions, but $C_{2m}={\bf Z}_2\times {\bf Z}_{4m}$ for even $m$ and ${\bf Z}_{8m}$ for odd $m$. $D_{2m-1}$ and $D_m$ for spin 1/2 fermion is ${\bf Z}_1$ and ${\bf Z}_2\times {\bf Z}_2$, respectively.

Efetov considered in ref.\cite{Efetov97} section 10.3,  one dimensional fermionic system using the partition function
\begin{equation}
Z=\int exp(-S[\Phi]) D\Phi,\nonumber
\end{equation}
where 
\begin{equation}
S[\Phi]=\int_{-\infty}^\infty tr[\frac{m}{2}(\frac{\partial\Phi(t)}{\partial t})^2+V(\Phi)]dt,\nonumber
\end{equation}
and $\Phi$ may be real symmetric, general Hermitian or composed of real quaternions. Corelation functions of interacting fermions were calculated.
The topological insulator predicted in \cite{AZ97} was applied to the integer quantum Hall effect and quantum spin Hall effect discovered by \cite{KM05b,KM05a} in the article of \cite{SRFL08, dNG15}. 

Kennedy and Zirnbauer\cite{KZ16} studied ${\bf Z}_2$ symmetric ground states of gapped superconductors using classes in the symmetric space. Interactions of electrons are taken into account, the ${\bf Z}_2$ symmetry can be broken to ${\bf Z}_8$ symmetry\cite{FK10}. The authors adopted the interaction similar to that of \cite{dAFF76},
\begin{equation}
\tilde H=t T+ w W_{tot}+v V_{tot}.\nonumber
\end{equation}

White\cite{White92} studied $S=1/2$ and $S=1$ antiferromagnetic one dimensional spin chain, using the density matrix formulation, including configurations of fractional $S=1/2$ spins at the ends of open $S=1$ chains.

The supersymmetry method based on using commuting and anticommuting variables for bosons and fermions, respectively, is a useful tool for studying disorder and chaos\cite{Efetov97}.

Supersymmetric extensions of Korteweg-de Vries (KdV) soliton equation \cite{MJD00}, Kadomtzev-Petviashvili (KP) equation \cite{MJD00} and the non-linear equation of Ablowitz-Kaup-Newell-Segur (AKNS) \cite{AKNS74} were studied by several authors \cite{MR85,CR86,MY93}.

Using the $D$ operator 
\begin{equation}
D_t f_n \cdot f_{n-1}=f_{n+1}f_{n-2} -f_n f_{n-1}, \quad D_x f\cdot g=\frac{\partial f}{\partial x}g-f\frac{\partial g}{\partial x}\nonumber
\end{equation}
the KdV equation is 
\begin{eqnarray}
&&\frac{\partial}{\partial t}u+\frac{\partial^3}{\partial x^3}u+6u\frac{\partial}{\partial x}u=0,\quad u=2\frac{\partial^2}{\partial x^2}(log f),\nonumber\\
&&(D_x^4-4D_xD_t) f\cdot f=0, 
\end{eqnarray}
and the KP equation is
\begin{eqnarray}
&&\frac{\partial}{\partial x_1}(\frac{\partial}{\partial x_3}u+\frac{\partial^3}{\partial x_1^3}u+6u\frac{\partial}{\partial x_1}u)+\frac{\partial^2}{\partial x_2^2}u=0, \nonumber\\
&&u=2\frac{\partial^2}{\partial x_1^2}\log\tau\nonumber\\
&& (D_{x_1}^4-4 D_{x_1}D_{x_3}+3D_{x_2}^2)\tau\cdot\tau=0
\end{eqnarray}
 
 The solution of AKNS equation in the inverse scatteing transform (IST) of Zakharov-Shabat\cite{ZS72,ZS73} is
\begin{equation}
q_t-\sqrt{-1}q_{xx}-2\sqrt{-1}q^2 q^*=0.
\nonumber
\end{equation}

In the ${\mathcal N}=2$ supersymmetric extension of KdV equation \cite{MY93}, the resolvent kernel $R=(-L+\zeta)^{-1}$ where $L=\partial_x^2-u(x)$, is expressed as
\begin{equation}
R(x,\zeta)=\lim_{x'\to x} R(x,x';\zeta)=\sum_n R_n[u]\zeta^{-n-1/2},
\nonumber
\end{equation}
and the residue or the coefficient of $\partial^{-1}$ becomes two times Gelfand-Dikii polynomials $R_n$,
\begin{equation}
res L^{n-1/2}=2R_n.
\nonumber
\end{equation}
The KdV differential equation is expressed by the differential equation of the Lax operator $L$, that satisfy
\begin{equation}
[L,L_+^{n+1/2}]=\frac{\partial}{\partial t_n}L\nonumber
\end{equation}
which is equivalent to 
\begin{equation}
\frac{\partial u}{\partial t_n}=4\partial_x R_{n+1}.
\nonumber
\end{equation}

Supersymmetric extension of the nonlinear AKNS equation was performed by introducing Lie 
superalgebra \cite{CR86}, in which the linear problem
\begin{eqnarray}
\Psi_x&=&(q E_2+i\lambda E_0+r E_1+\epsilon E_3+\beta E_4)\Psi,\nonumber\\
\psi_t&=&V\psi,
\end{eqnarray}
where $\epsilon, \beta$ are anticommuting variable, $r,q$ are commuting fields, 
\begin{equation}
V=\left(\begin{array}{ccc}
A&C&\alpha\\
B&-A&\rho\\
\rho&-\alpha&0\end{array}\right)\nonumber
\end{equation}
is the temporal evolution operator, $A,C$ are commuting and $\alpha,\rho$ anticommuting values, are solved. $\lambda$ is the eigenvalue of the problem and $\psi$ is the temporal evolution.

The state vectors $^t(B_j, C_j, \rho_j, \alpha_j)$ satisfy with Lie superrecursion operator $\mathcal L$ as
\begin{equation}
2\sqrt{-1}\,  ^t(B_{j+1},C_{j+1},\rho_{j+1},\alpha_{j+1})={\mathcal L}\, ^t(B_j,C_j,\rho_j,\alpha_j).\nonumber
\end{equation}
The $E_i$'s satisfy commutation rules
\begin{eqnarray}
&&[E_0,E_1]=-2E_2, \quad [E_0,E_2]=2E_1,\quad [E_1,E_2]=E_0, \nonumber\\
&& [E_0,E_3]=E_3, \quad [E_0,E_4]=-E_4,\nonumber\\
&&[E_1, E_3]=0, \quad [E_1, E_4]=E_3,\quad [E_2,E_4]=0,\nonumber\\
&&\{E_3, E_4\}=E_0, \quad \{E_3,E_3\}=-2E_1,\nonumber\\
&&  \{E_4,E_4\}= 2E_2.
\end{eqnarray} 

In the case of supersymmetric KP equation\cite{MR85}, pseudo-differential operator $L=\partial+\sum_{i=1}^\infty u_i \partial^{-i}$ and the differential equation
\begin{equation}
\partial_i L=[L_+^i,L], \quad \partial_i=\frac{\partial}{\partial t_i},
\nonumber
\end{equation}
where $L_+$ is the differential part of the operator $L$, was solved. Here even-odd pairs of space variables $(x,\xi)$ and even-odd times $(\tau_1,t_2,\tau_3,t_4,\cdots)$ were adopted.
\section{Symmetries in propagation of solitary waves in matters}
Although magnetic monopoles are not detected, they can be regarded as 3D solitons, and supersymmetry have relevance to solitons.
Time reversal symmetry based nonlinear elastic wave spectroscopy, in which one optimizes the convolution of the scattered wave from defects in materials and its time reversed wave show peaks was an effective method for non-destructive testing (NDT)\cite{BLEFF16}.

Getting physical information of nonlinear acoustic equations was performed by using the Lie symmetry of the Khokhlov-Zabolotskaya(KhZa) equation which uses the front form coordinate $\theta=\omega(t-x/c_0)$ and a special coordinate $\eta(r,z,u)$\cite{DosSantos04,DSBM04}. 
Details on possibility of detecting gravitational effects using AI technique is discussed in \cite{FDS20b}.

\subsection{Conformality in quaternion projective space}
In \cite{JLW87}, orthogonal projection of the Dirac operator $Q_+=P_- Q P_+$, where $P_\pm=\frac{1\pm \Gamma}{2}$ and index  $i(Q_+)$ for Wess-Zumino model defined by the superpotential $V(\varphi)$ where $\varphi$ is the holomorphic function ($\varphi\in{\bf C}$) becomes Atiyah-Singer index
\begin{equation}
i(Q_+)=dim Ker Q_+ -dim Ker Q_-=n_+-n_-
\end{equation} 
was verified.
The grading $\Gamma$ is defined by fermionic particle number $N_f$ as $\Gamma=(-1)^{N_f}$.

In TR-NEWS experiment, propagation of a soliton is restricted to a $2D$ plane, and the support of $\varphi$ has rectangular boundary. An index theorem for Dirac operators in a bounded region was formulated by Atiyah, Patodi and Singer\cite{APS75}, and discussed by Witten\cite{Witten15} and Yu et al\cite{YWX17}.  ${\bf Z}_2$ topological order and quantum spin Hall effect, which are related to boundary conditions of conformal field theory was investigated in 2005 by Kane and Mele\cite{KM05b, KM05a}.  

Analtical extensions of differentiable functions defined in closed sets was discussed in 1933 by Whitney\cite{BT82}, refined by Seeley\cite{Seeley64}, and boundary conditions folllowing the heat equation was studied by Atiyah, Bott and Patodi\cite{ABP73}. A review of differential calculus of functions that map open interval $I$ to $n D$ real functions $f=(f_1,\cdots, f_n)\in C^k$ by Taylor expansions
\begin{eqnarray}
&&f(x+y)=\sum_0^{k-1} f^{(j)}(x;y,\cdots y)/j!\nonumber\\
&&+\int_0^1f^{(k)}(x+ty; y,\cdots ,y)(1-t)^{k-1} dt/(k-1)!, \nonumber\\
\end{eqnarray}
where $[x,x+y]\in I$ and their Fourier-Laplace transform were shown by Hoermander\cite{Hoermander83}. A good review on Atiyah-Singer's index theorem\cite{AS63} and Atiyah-Bott-Shapiro's Clifford modules\cite{ABS64} are written in the book of Hirzebruch\cite{Hirzebruch78}. Stability of index formulas and boundary problems for elliptic differential operators are explained in \cite{Hoermander87}.

In application of mathematical theories to speach recognitions or image recognitions, recurrent neural networks are commonly used\cite{Aggarwal18}. Parcollet et al.\cite{PRMLTDMB19} applied quaternion neural networks to image recognitions. 
Although the subject is not directly related to the supersymmetry, in the analysis of propagation of solitons which are expressed by conformal wave function on rectangular $2D$ plane, edge effects of solitoic wave should be taken into account, where Atiyah-Padobi-Singer index could play a role in the analysis in quaternion projective space\cite{FDS20b}.
In the analysis of soliton spectroscopy, the projective quaternions space $q_1$ and $q_2$ that satisfy $q_1h-hq_2=0$ for a quaternion $h$ are equivalent\cite{MGS14}.

Applications of memoducers to nondestructive testing is recently reviewed by Dos Santos \cite{DosSantos20}.
Analysis of noise in zero-mode solutions of circuits was performed in\cite{DSP07}. Measurement of effects of instantons which are related to the supersymmetry and gravitational forces due to changes of metrics are future problems.

In the analysis of spectroscopy of QED and QCD, the complex projective space $P_m$ in which for $(z^0,z^1,\cdots,z^m)$ and $z^i\ne 0$, only local coordinates $_i\zeta^j=z^k/z^i$ $(k\ne i)$ can be used\cite{Chern67}. 
  
\section{Discussion and Conclusion}
Supersymmetry in hadron dynamics is important to fix poperties of Higgs bosons, they define masses of quarks, leptons, vector mesons and hadrons.   In nature spersymmetry is broken, and super partners are usually undetected. In high energy hadron dynamics, parton models which are asymptotically free in high energy have problems in the infrared region. Lattite QCD simulation suggest that quarks are confined and the running couplings between quarks is infrared stable. 

Unification of the gravitational field and the standard model through string theory has not been successful\cite{Polyakov87}. It may be due to difficulty of spontaneous symmetry breaking, as manifested by the presence of Higgs bosons. Magnetic monopoles, which are solitons in $3D$ space are not conformal \cite{Ebert89,Dunajski10}. Even when the potential is singular, the field propagators are not necessarily singular.

In analyzing hadron dynamics, duality of conformal field theory and holographic treatement in AdS space was proposed. Since quarks has masses, QCD is not conformal, but energy dependence of dynamics suggest that there are conformal windows which patch from low energy regions to high energy regions. 

Local conformal structure seems to apply not only to hadron dynamics, but also dynamics of solitons, which have connections to solid state physics.

For characterization of the manifold, conformality played an important role, since $\frac{\partial f}{\partial \bar z}=0$ reduces the freedom of complex functions $f(x,y)=f(z)$.

In $2D$ field theory, Mermin-Wegner-Hohenberg theory\cite{MW66, Hohenberg67} says that in 2D, ferromagnetism or antiferromagnetism are absent in isotropic Heisenberg model. Coleman\cite{Coleman85} also claimed that in $2D$ electromagnetic field theory spontaneous symmetry breaking is absent. The success of LFHQCD may be due to the choice of a parameter of Schroedinger equation $\zeta^2=x(1-x){\bf k}_\perp$, which is the $2D$ Fourier transform of ${\mathcal M}^2$\cite{Brodsky19,Sandapen20}. Spontaneous symmetry breaking or Higgs effects do not appear in LFHQCD. Chua's observation of currents of memristor\cite{Chua71,Chua18} suggests validity of patched LFHQCD solutions of different $\tau$.

Recently LFHQCD model succeded to explain polarized and unpolarized quark distributions in the proton \cite{dTLSDBD18,LSdTDBD20}.

Braam and van Baal\cite{BvB89} showed that mapping from anti-self dual connections on four torus $(S^1)^4$ onto
anti-self dual connections on the dual torus which is called the Nahm transformation induces a map between the relevant instanton modulei spaces\cite{AHDM77,BP13}. They considered quaternionic structures on spinors on 4-manifold, and as a special choice ordinary instantons on $S^4$. van Baal and van den Heuvel\cite{vBvdH94} extended the idea to the GZ theory of SU(2) on ${\bf R}^3$ and constructed the theory of sphalerons analogous to instantons. Since ${\bf R}^4$ is conformally equivalent to $S^3\times {\bf R}$, gauge fixing was done on $S^3\times {\bf R}$ manifold. By connecting $t=+\infty$ and $t=-\infty$ in $\bf R$, one can consider effects of instanton.

I changed the global boundary condition of ${\bf R}$ from $t=-\infty$ to $t=\infty$ to local periodicity one and considered $S^3\times S^1$ which is isomorphic to complex projective space ${\bf CP}^3$\cite{Kodaira92}.
Clifford algebra on the $S^1\times S^3$ can be expressed by quaternions\cite{AHDM77,SF20, SF20a}.

In the twistor formulation of Dunajski\cite{Dunajski10}, projection of ${\bf C}^2\to {\bf CP}^1$ was considered and stability of exponentiatial of infinitesimal deformation proven by Kodaira\cite{Kodaira63} was applied.  On a compact surface $S$, Kodaira \cite{Kodaira66} defined an automorphism $g$ of a domain $W$ given by $(z_1,z_2)={\bf C}^2-(0,0)\sim {\bf CP}^1$,
\begin{equation}
g: (z_1,z_2)\to (z_1',z_2')=(s z_1+\lambda z_2^m, t z_2)
\nonumber
\end{equation}
where $m$ is a positive integer and $s,t,\lambda$ are constants such that
\begin{equation}
0<|s|\leq|t|<1, \quad (t^m-s)\lambda=0.
\nonumber
\end{equation}
and for an infinite cyclic group ${\bf Z}$, the quotient surface $W/Z$ was homeomorphc to ${ S}^1\times { S}^3$. It was shown that the cohomology groups $H^\nu({S}^1\times {S}^3, {\bf C}(F_\mu))$, $\nu=0,1,2,\cdots$vanish for $\mu\ne 1$, which means that there is a curve on the patched complex areas on the surface $S$.

In the analysis of solitons propagating on $2D$ planes, quaternion projective space may be useful, by identifying real quaternions $p_1$ and $p_2$, if there exists a quaternion ${\bf H}\setminus \{0_{\bf H}\}$ such that $hp_1=p_2h$ for reducing parameters\cite{MGS14}. In literatures\cite{Swan62}, projective quaternion spaces were defined on tangent bundle of ${S}^4$ manifold. But they can be defined on $2D$ plane bundle parametrized by $\tau^\pm=t\pm z/c$\cite{FDS20b,SF20a}.

As a model of Higgs boson based on $SU(4)/Sp(4)$ was also discussed in \cite{BGR14, ACCDLCS17}. Experiments of enhancement of Higgs boson decay into $\gamma\gamma$ due to WZW term via the technicolor was investigated. 

As an extension of the Maxwell-Gravitation model, complex octonion model\cite{Weng16} in which a quarternion ${\bf H}_e$ is used for the electromagnetic field and the second quaternion ${\bf H}_g$ is used for gravitational field and complex octonion field
\begin{eqnarray}
&&{\bf H}(h^\alpha)=\sqrt{-1}h^0{\bf i}_0+h^r{\bf i}_r+\sqrt{-1}h^4{\bf i}_4+h^{4+r}{\bf i}_{4+r},\nonumber\\
&& (r=1,2,3)
\nonumber
\end{eqnarray}
was proposed. In the curved space metric was defined as $g_{\bar\alpha\beta}\bar{du^\alpha}du^\beta$, depending on the coordinates of quaternions. It is not clear whether the affine connection of complex octonions can express correct curvature in Maxwell-Gravitation models. Metric tensors $g_{\bar\alpha\beta}$ should satisfy
\begin{equation}
g_{\bar\alpha\beta}(x)=\eta_{ab} e^a_{\bar\alpha}(x)e^b_{\beta}(x)=e^a_{\bar\alpha} e_{b\beta}
\nonumber
\end{equation}
where $\eta$ is the Minkowski metric, and $e_{b\beta}=\eta_{ab}e^b_\beta$.  Whether such a metric can be defined by quaternion basis vectors instead of usual spinors is not obvious.

 In a $SU(2)$ spharelon model of Yang-Mills equation\cite{vBHD92}, the Riemannian metric and the curvature of $S^3\times {\bf R}$ manifold were calculated.

In LFHQCD\cite{dTDB15}, the metric is contained in the dilaton background field factor and absorbed in the Rarita-Schwinger spinor.

Constraints on supersymmetry breaking was studied by Witten \cite{Witten82b} and spontaneous symmetry breaking of $SU(3)_L\times SU(3)_R$ to the diagonal $SU(3)$ was discussed in \cite{Witten88}. Bases of these theories seems to exist in Wess-Zumino's theory\cite{WZ74}, as written in my review on supersymmetry.
There are no BRST ghosts in the holographic lightfront QCD (HLFQCD) approach of Brodsky et al. and K-theory approach of Connes.

From the 3 dimensional linear subspace of ${\bf R}^3$, one can define the sphere $S^2$ and mapping $S^2\to {\bf P}^2$ i.e.projective subspace in manifold $S^3$.  Quaternion projective space was considered to reduce parameters to fit physical data.
I would like to point out that supergroup functions in $2D$ using projective quaternion space can be used in physics and engineering which was be discussed elsewhere\cite{FDS20a,FDS20b}. It is my impression that the conformality of holonomic functions on projective space which allow patching local coordinates to global one is crucial, and the  Hopf manifold proposed by Chern\cite{Chern67} can be used to understand dynamics of nature.

The supersymmetry was applied to hadron spectrum region, solid state phisics region and Physics of universe including the dark matter.
I presented a biased view of supersymmetry based on the theory of Fubini and his collaborators, and Brodsky and his collaborators related quark model to particle physics and supersymmetry based on the theory of Efetov and his collaborators related to low dimensional spin dynamics and solitons. 

There are many good references on supersymmetry, and I apologize to authors whom I did not mention in this review. From mathematical point of view of supersymmetry, DeWitt's book \cite{DeWitt92}, Freed's book\cite{Freed99}, and a concise encyclopedia of supersymmetry edited by Duplij, Siegel and Bagger \cite{DSB04}  and references therein are helpful.

\vskip 0.5 true cm

{\bf Aknowledgement}
I thank Stanley Brodsky for sending information of works in his group, Guy de T\'eramond and  Serge Dos Santos for helpful discussions, late Professor Hideo Nakajima for collaboration on lattice simulations, physics research organizations in France, Germany for giving me oppotunities to study symmetry between 1972 and 1989, Dr. Claude Amsler for attracting my attention to \cite{CMS20} and Professor Steven Duplij of Kharkov University in Ukraine for sending me the book of encyclopedia of supersymmetry\cite{DSB04} in 2005.
 
Thanks are also due to libraries of Tokyo Institute of Technology and the library of the institute of mathematical science of the University of Tokyo for allowing consulting references.

\vspace{6pt} 






\end{document}